\documentclass[aps,prb,twocolumn,superscriptaddress,showpacs]{revtex4-1}

\usepackage{graphicx} % Include figure files
\usepackage{amssymb}
\usepackage{color}
\usepackage{amsmath}
\usepackage{hyphenat}
\usepackage{hyperref}

\begin{document}

\title{Experimental and theoretical electronic structure of quinacridone }

%% Notice placement of commas and superscripts and use of &
%% in the author list

%\author{Daniel L\"uftner$^{1}$}
%\author{Sivan Refaely-Abramson$^{2}$}
%\author{Michael Pachler$^{1,3}$}
%\author{Roland Resel$^{3}$}
%\author{Michael G. Ramsey$^{1}$}
%\author{Leeor Kronik$^{2}$}
%\author{Peter Puschnig$^{1}$}\email{peter.puschnig@uni-graz.at}

%\affiliation{$^{1}$Institute of Physics, University of Graz, NAWI Graz, Universit\"atsplatz 5, 8010 Graz, Austria}
%\affiliation{$^{2}$Department of Materials and Interfaces, Weizmann Institute of Science, Rehovoth 76100, Israel}
%\affiliation{$^{3}$Institute of Solid State Physics, Graz University of Technology, NAWI Graz, Petersgasse 16, 8010 Graz, Austria}

\author{Daniel L\"uftner}
\affiliation{Institute of Physics, University of Graz, NAWI Graz, Universit\"atsplatz 5, 8010 Graz, Austria}
\author{Sivan Refaely-Abramson}
\affiliation{Department of Materials and Interfaces, Weizmann Institute of Science, Rehovoth 76100, Israel}
\author{Michael Pachler}
\affiliation{Institute of Physics, University of Graz, NAWI Graz, Universit\"atsplatz 5, 8010 Graz, Austria}
\affiliation{Institute of Solid State Physics, Graz University of Technology, NAWI Graz, Petersgasse 16, 8010 Graz, Austria}
\author{Roland Resel}
\affiliation{Institute of Solid State Physics, Graz University of Technology, NAWI Graz, Petersgasse 16, 8010 Graz, Austria}
\author{Michael G. Ramsey}
\affiliation{Institute of Physics, University of Graz, NAWI Graz, Universit\"atsplatz 5, 8010 Graz, Austria}
\author{Leeor Kronik}
\affiliation{Department of Materials and Interfaces, Weizmann Institute of Science, Rehovoth 76100, Israel}
\author{Peter Puschnig}\email{peter.puschnig@uni-graz.at}
\affiliation{Institute of Physics, University of Graz, NAWI Graz, Universit\"atsplatz 5, 8010 Graz, Austria}

%\affiliation{$^{1}$Institute of Physics, University of Graz, NAWI Graz, Universit\"atsplatz 5, 8010 Graz, Austria}
%\affiliation{$^{2}$Department of Materials and Interfaces, Weizmann Institute of Science, Rehovoth 76100, Israel}
%\affiliation{$^{3}$Institute of Solid State Physics, Graz University of Technology, NAWI Graz, Petersgasse 16, 8010 Graz, Austria}

\date{\today}

\begin{abstract}
The energy positions of frontier orbitals in organic electronic materials are often studied experimentally by (inverse) photoemission spectroscopy and theoretically within density functional theory. However, standard exchange-correlation functionals often result in too small fundamental gaps, may lead to wrong orbital energy ordering, and do not capture polarization-induced gap renormalization.
Here, we examine these issues and a strategy for overcoming them by studying the gas phase and bulk electronic structure of the organic molecule quinacridone (5Q), a promising material with many interesting properties for organic devices. 
Experimentally, we perform angle-resolved photoemission spectroscopy (ARUPS) on thin films of the crystalline $\beta$-phase of 5Q. Theoretically, we employ an optimally-tuned range-separated hybrid functional (OT-RSH) within density functional theory.
For the gas phase molecule, our OT-RSH result for the ionization potential (IP) represents a substantial improvement over the semi-local PBE and the PBE0 hybrid functional results, producing an IP in quantitative agreement with experiment. 
For the bulk crystal, we take into account the correct screening in the bulk, using the recently developed OT-SRSH approach, while retaining the optimally-tuned parameters for the range-separation and the short-range Fock exchange. This leads to a band gap narrowing due to polarization effects and results in a valence band spectrum in excellent agreement with  experimental ARUPS data, with respect to both peak positions and heights. Finally, full-frequency $G_0W_0$ results based on a hybrid functional starting point are shown to agree with the OT-SRSH approach, improving substantially on the PBE-starting point.

\end{abstract}
\maketitle
\section{Introduction}

Organic semiconducting devices are extensively studied as they have the advantage, compared to their Si-based counterparts, of being flexible, cheap, light-weighted and processable at low temperatures.\cite{Klauk2006,Yamashita2009} Recently, the organic pigment  5,12-dihydro-quino[2,3-b]acridine-7,14-dione (quinacridone), C$_{20}$H$_{12}$N$_2$O$_2$, a derivative of pentacene, has attracted considerable interest as an active material in organic electronics,\cite{Wang2007b,Chen2010,Glowacki2012,Wagner2014,Song2014} e.g. due to its remarkable air-stability, high photogeneration efficiency, electrochemical stability, and high fluorescence lifetime in solution.

Quinacridone (5Q) differs from pentacene by the presence of functional N-H and C-O groups at phenyl rings 2 and 4 (see Fig.~1a). On the one hand, these polar groups cause the formation of intermolecular hydrogen bonds. These are responsible for 5Q's ability to form self-assembled, supramolecular structures and allow for the remarkable air-stability of 5Q devices reported in the literature.\cite{Trixler2007} However, these functional groups also break the conjugation of the molecule and thereby significantly change the energetic positions and spatial shapes of the frontier orbitals. The latter are crucial for the performance of organic semiconducting devices because they determine electron removal and insertion energies and therefore the band gap and level alignment. As such, they are extensively studied in the context of organic molecular systems and organic/inorganic interfaces.\cite{Kahn2003,Ueno2008,Puschnig2011,Koch2013} Such investigations strongly benefit from accurate and computationally inexpensive theoretical models, which are often necessary for a better interpretation of experimental results.\cite{Kronik2013} 

An efficient work horse for first principles calculations of the electronic structure is density functional theory (DFT),\cite{ParrYang1989,Gross1995} usually within the Kohn-Sham (KS) framework.\cite{Hohenberg64,Kohn1965} In this scheme, the original many-electron problem is mapped uniquely into a fictitious noninteracting electron system, yielding the same electron density. This mapping leads to effective single-particle equations that provide a significant conceptual and computational simplification of the original many-electron problem. However, due to the fictitious nature of the noninteracting electrons, the correspondence of KS eigenvalues with ionization energies measured in an experiment is not straightforward.\cite{Sham66-2} One exception is the highest occupied orbital whose energy can be rigorously identified with the ionization potential (IP) of the neutral system, if the \emph{exact} exchange-correlation (xc) functional is used.\cite{Perdew1982,Levy1984,Almbladh1985,Perdew1997} 
In general, KS results based on \emph{approximate} xc expressions, e.g. local or semi-local functionals,  may suffer from pronounced self-interaction errors (SIE) and a lack of the derivative discontinuity in the xc potential, and therefore are not expected to agree with experimental findings obtained, e.g. from photoemission spectroscopy.\cite{Kummel2008} These conceptual problems may, at least partly, be cured by introducing the generalized Kohn-Sham (GKS) scheme,\cite{Seidl1996} and considering hybrid functionals with a fraction of exact exchange either globally or in a range-separated manner.\cite{Kummel2008} However, band gaps, IPs, electron affinities (EAs), and the orbital order obtained within such hybrid functional may still be in error when compared to experiment.\cite{Kronik2012,Kronik2014}

% with 

A promising strategy to improve the agreement with experiment is the more recent class of range-separated hybrid (RSH) DFT functionals,\cite{Iikura2001,Leininger1997} where the interelectron Coulomb repulsion term is separated into long-range (LR) and short-range (SR) components. The LR term is mapped using Hartree-Fock theory, thereby establishing the correct asymptotic potential. The SR term is mapped using a (semi-)local KS functional or a conventional hybrid functional, maintaining the compatibility between the exchange and correlation expressions. In this approach, one still needs to determine the range-separation parameter,\cite{Baer2005} as a universal value usually leads to energy levels that, although greatly improve the accuracy of standard hybrids, are still not at the desired accuracy level.\cite{Baer2010,Kronik2012} This can be improved by using optimally tuned RSH (OT-RSH) functionals,\cite{Salzner2009,Kronik2012} where the range-separation parameter is tuned for each system
such that physically motivated tuning conditions are fulfilled
without introducing any empirical parameters. DFT calculations using OT-RSH functionals have been shown to provide an accurate, non-empirical description of band gaps, IPs and EAs for a variety of systems, among them atoms, molecules and polymers \cite{Stein2010,Foster2012,Sun2014,Korzdorfer2011a} as well as larger gas phase organic molecules relevant for organic semiconducting devices. \cite{Refaely-Abramson2011,Phillips2014} In addition, it has been demonstrated that the description of deeper lying occupied states of an isolated molecule can be considerably improved within the OT-RSH approach, if an additional degree of freedom is introduced, which allows the modification of exact exchange in the short range. \cite{Refaely-Abramson2012,Egger2014} In particular, it was shown to well-describe the outer-valence spectra of several organic molecules that exhibit a mixture of localized ($\sigma$) and delocalized ($\pi$) states, \cite{Refaely-Abramson2012} a challenging situation where the difference in self interaction error (SIE) for different orbital types can lead to the wrong description of orbital-ordering with standard DFT methods. \cite{Marom2009,Korzdorfer2009,Korzdorfer2010,Korzdorfer2010a,Rissner2011}

Beyond gas-phase molecules, band gaps of various organic molecular \emph{crystals} have also been recently successfully described with an OT screened-RSH (OT-SRSH) functional.\cite{Refaely-Abramson2013} This was achieved by including a new constraint for the asymptotic behavior of the exchange correlation potential, thereby taking into account effects arising from polarization-induced band renormalization. \cite{Sato1981,Neaton2006,Sharifzadeh2012} An open question, however, is whether the OT-SRSH functional is capable of accurately predicting not only band gaps, but the entire outer valence spectrum of molecular crystals. In  particular, it is interesting to examine the OT-SRSH accuracy when the crystal is comprised of more complex organic molecules, such as 5Q, that are characterized by a mix of localized ($\sigma$) and delocalized ($\pi$) states as frontier orbitals, where different SIEs are expected.  It has not yet been investigated whether the OT-SRSH approach can accurately deal with self-interaction problems and at the same time cope with polarization effects arising from the crystalline environment. Capturing both is necessary for an overall good description of the electronic structure. For this purpose, 5Q turns out to be an ideal test candidate.

In this article, we report a combined experimental and theoretical study of the electronic structure of 5Q, which answers the above question. We performed angle resolved UPS experiments taken on well-ordered films of 5Q in the $\beta$-phase,\cite{Paulus2007,Panina2007} and provide a detailed theoretical assignment of the various peak positions, using the OT-SRSH method. To this end, we first investigated the isolated 5Q molecule, by performing OT-RSH calculations. This yielded an IP in excellent agreement with literature data from gas phase ultraviolet photoemission spectroscopy (UPS). We then calculated the electronic structure of the bulk $\beta$-phase, taking into account the correct asymptotic behavior in the OT-SRSH approach by computing the dielectric constant of the bulk crystal within the random phase approximation. To gain a better understanding of our results, we also performed many-body perturbation theory calculations, within the G$_{0}$W$_{0}$ approximation \cite{Hedin65,Hybertson86} using various DFT starting points. We obtained excellent agreement with experimental results for both the OT-SRSH and G$_{0}$W$_{0}$ calculations, the latter based on a DFT starting point obtained from a conventional hybrid functional.

\section{Experiment}

5Q films were grown in-situ in ultra-high vacuum (UHV) at room temperature on an atomically clean and ordered Cu(110) substrate. The Cu surface was prepared by repeated cycles of Ar$^+$ -ion bombardment and annealing at 800 K. A 260 {\AA} thick 5Q film was deposited {\it in situ} from a thoroughly degassed evaporator, such that the pressure in the system remained at the $10^{-10}$ mbar range during film growth. The nominal growth rate was 2 {\AA} min$^{-1}$, as monitored by a quartz microbalance assuming a density of 1.47 g cm$^{-3}$. 

The electronic structure has been characterized in-situ with UPS. Angle-resolved photoemission (ARUPS) experiments were performed using a VG ADES 400 spectrometer described elsewhere.\cite{Oehzelt2007} The ADES system was equipped with a noble gas discharge lamp (unpolarized helium I radiation, $h\nu = 21.2$ eV) and a movable electron energy analyzer, allowing angle resolved ultraviolet photoelectron spectroscopy in the specular plane, with an angular resolution of ±1$^\circ$ and a total energy resolution of 150 meV at room temperature. A photon incidence angle of $\alpha$ = 60$^\circ$ was used. After the ARUPS measurements, the sample was removed from vacuum for geometrical structure investigations, ex-situ, using x-ray diffraction (XRD) with both $\Theta/2\Theta$ scans and pole figure analysis (Philips X'PERT four circle texture goniometer).\cite{Novak2011} 

The XRD data revealed the $\beta$-phase 5Q polymorph,\cite{Paulus2007,Panina2007} oriented with its ($\overline{1}12$) plane (blue line in Fig.~1b) parallel to the substrate, with four equivalent domains. As illustrated in Fig.~1, in any one domain of $\beta$-5Q the axes of the molecules are almost parallel to the substrate surface ($\approx$  7$^\circ$). Due to the two molecules in the unit cell having their aromatic planes at $\approx$ 70$^\circ$ to each other, and the multiplicity of domains, only small angular variations were observed in the angle-resolved UPS. Orbital tomography predictions of the angular distribution \cite{Puschnig2011} suggested that both $\pi$ and $\sigma$ orbital emissions contribute to the spectra. Thus the experimental spectra may be safely related to the calculated density of states.

\begin{figure*}[!htb]
	\includegraphics[width=16cm]{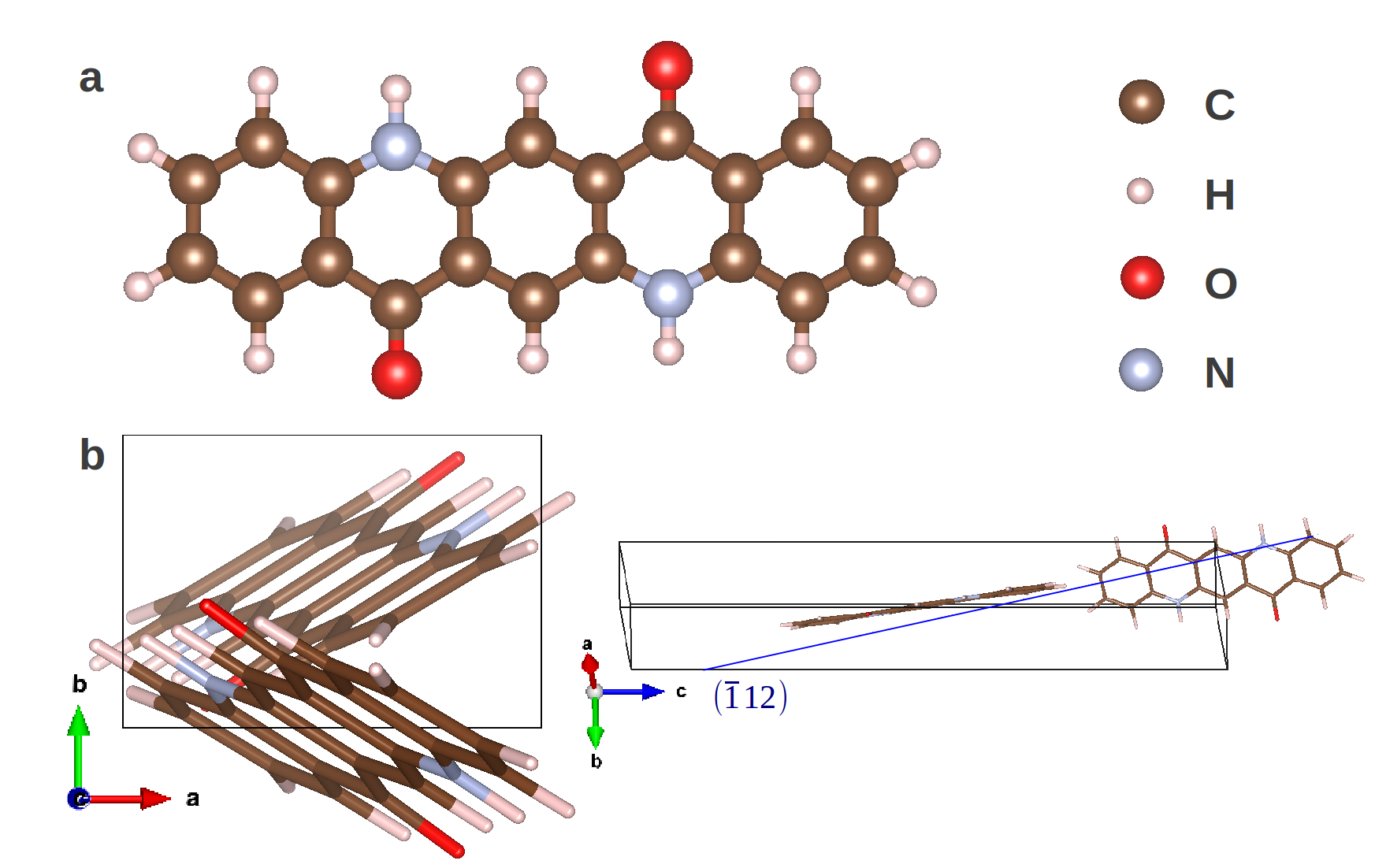}
	\caption{\label{Fig1} (a) Schematic view of the 5Q molecule. (b) Alignment of the 5Q molecules in the $\beta$-crystalline structure, with the ($\overline{1}12$) plane indicated as a blue line.}
\end{figure*}  

\section{Theory}

The electronic structure of both the isolated 5Q molecule and its $\beta$-crystal structure has been calculated using two different types of electronic structure approaches. The first type is within the framework of DFT, where the exchange-correlation potential is approximated in several different ways: using the generalized gradient approximation (GGA), as parametrized by Perdew, Burke and Ernzerhof (PBE); \cite{Perdew1996} The global hybrid PBE0;\cite{Adamo1999} The short-range hybrid of Heyd, Scuseria and Ernzerhof (HSE); \cite{Heyd2006} And finally the optimally-tuned range-separated hybrid (OT-RSH) functional for isolated molecule calculations \cite{Refaely-Abramson2011} and the recently proposed optimally-tuned screened range separated hybrid (OT-SRSH) \cite{Refaely-Abramson2013} functional for the molecular solid, respectively. The second type of calculation is based on many-body perturbation theory, within the GW approximation, \cite{Hedin65} calculated as perturbative "one-shot" G$_{0}$W$_{0}$ \cite{Hybertson86} corrections to DFT-based eigenvalue spectra, from either PBE or HSE calculations. While the G$_{0}$W$_{0}$ approach has been often applied to molecules and molecular solids and is well represented in the literature (see, e.g., Ref. \cite{Tiago2003,Tiago2006,Neaton2006,Garcia-Lastra2009,Marom2011,Blase2011,Sharifzadeh2012,Puschnig2012,Cudazzo2012,Faber2014,Droghetti2014}),  OT-SRSH based calculations are rather new and therefore we provide a concise overview of the basic ideas of that approach. For overviews from different perspectives, the reader is referred to Refs. \onlinecite{Baer2010,Kronik2012,Korzdorfer2014,Faber2014,Autschbach2014} and specifically for studies of organic molecular crystals to Ref \onlinecite{Refaely-Abramson2013}.

\subsection{Optimally-tuned range-separated hybrids}

The starting point of the range-separated hybrid (RSH) concept, which is couched within the GKS formalism, \cite{Seidl1996,Kronik2012} consists of a partition of the Coulomb interaction as: \cite{Yanai2004,Srebro2012,Srebro2012a}

\begin{equation}\label{eq:1}
\frac{1}{r} =  \frac{\alpha + \beta \, \textrm{erf}(\gamma r)}{r} + \frac{1-[\alpha+\beta \, \textrm{erf}(\gamma r)]}{r} 
\end{equation}
Here, $r$ is the inter-electron coordinate, $\textrm{erf}$ is the error function and $ \alpha $, $ \beta $, and $ \gamma $ are parameters, which in principle may be freely chosen or determined empirically.\cite{Yanai2004} The two parts of the split Coulomb operator are treated differently when computing the exchange interaction. While the first term is treated within Hartree-Fock (HF) theory, the second one is treated within a standard semi-local (sl) approximation. \cite{Leininger1997} The parameter $ \gamma $ is the so-called range-separation parameter. It controls which of the two terms dominates at a given range. It has been repeatedly shown that one uniform value for $ \gamma $ is not sufficient in every case, and that in fact $\gamma$ can be strongly system- and size-dependent. \cite{Refaely-Abramson2011,Stein2010,Korzdorfer2011a,Kronik2012,Tamblyn2014} Therefore, we prefer to determine it separately for each system by fulfilling a non-empirical condition. This is the point where the optimal-tuning comes into play: $\gamma$ is chosen such that the difference between the energy of the HOMO level and the IP is minimized, i.e., we make use of the ionization-potential theorem.\cite{Perdew1982,Levy1984,Almbladh1985,Perdew1997} Practically, this is achieved  by minimizing the following target function:\cite{Stein2010}
\begin{equation}\label{eq:2}
J^2(\gamma;\alpha) = \sum \limits_{i} [\epsilon_{H(N+i)}^{\gamma,\alpha}+ IP^{\gamma,\alpha}(N+i)]^2.
\end{equation}
Here, the \( \epsilon_{H(N+i)}^{\gamma,\alpha} \) are the HOMOs of the \( (N + i) \) electron molecular systems and $i$ is an integer number. The \( IP^{\gamma,\alpha}(N+i) \) are the corresponding ionization potentials, which are calculated from the total energy difference between the  N+i electron and the  N+i-1 electron system. For the moment, $\alpha$ remains as a free parameter and the tuning is performed for each choice of $\alpha$ separately, yielding different optimal $\gamma$ values that minimize \( J^2(\gamma;\alpha) \). Including \( i = 0 \) in the sum of Eq.~(\ref{eq:2}) equalizes the HOMO of the neutral system with the IP, while for \( i = 1 \) the IP of the anionic system, i.e., the electron affinity of the neutral system is considered, such that its difference from the LUMO eigenvalue is minimized.\cite{Kronik2012} By extension of the sum to negative values of $i$, states beneath the HOMO are also accounted for. This has been shown to assist in obtaining optimal tuning if the corresponding orbitals are strongly localized.\cite{Refaely-Abramson2012}   

The remaining parameters $\alpha$ and $\beta$, appearing in Eq.~(\ref{eq:1}), control the behavior of the Fock term at its limits. It tends to \( \frac{\alpha}{r} \) when $r \rightarrow 0$ and  to \( \frac{\alpha +\beta}{r} \) when $r \rightarrow \infty$. The asymptotic behavior of the xc functional for $r \rightarrow \infty$ has been shown to be crucial for obtaining accurate gaps between the highest occupied molecular orbital (HOMO) and the lowest unoccupied molecular orbital (LUMO) and introduces a second constraint on the parameters. For an isolated molecule, the correct  \( \frac{1}{r} \) asymptotic behavior is thus achieved by enforcing \( \alpha + \beta = 1 \). \cite{Rohrdanz2008} As a consequence, the semi-local contribution in the long range is set to zero and $\alpha$ now controls the amount of nonlocal Fock exchange in the short range. This can be seen in the expression for the exchange-correlation energy of the RSH functional, obtain with this constraint:\cite{Refaely-Abramson2012}
\begin{equation}\label{eq:3}
E_{xc} = (1- \alpha) E_{sl,x}^{SR} + \alpha E_{HF,x}^{SR} + E_{HF,x}^{LR} + E_{sl,c},
\end{equation}
where $sl,x$ and $sl,c$ denote semi-local exchange and correlation, respectively, and $HF,x$ denotes non-local Fock exchange. When moving from an isolated molecule to an organic crystal, the asymptotic behavior of the xc potential must take into account dielectric screening effects in the bulk. Thus for $r \rightarrow \infty$, the correct limit should be \( \frac{1}{\varepsilon r}  \), where $ \varepsilon$ is the scalar dielectric constant. This can be achieved by choosing $\alpha$ and $\beta$ such that the condition \( \alpha +\ \beta = \frac{1}{\varepsilon} \) is fulfilled. Note that a gas phase calculation may be seen as a special case of this constraint with $\varepsilon$ being equal to one. For the case $\varepsilon \neq 1$, the expression for the exchange-correlation energy becomes:
\begin{multline}\label{eq:4}
E_{xc} = (1- \alpha) E_{sl,x}^{SR} + \alpha E_{HF,x}^{SR} + (1 - \frac{1}{\varepsilon}) E_{sl,x}^{LR}+ \\ +  \frac{1}{\varepsilon} E_{HF,x}^{LR} + E_{sl,c}.
\end{multline}
Comparing Eqs.~(\ref{eq:3}) and (\ref{eq:4}), clearly the new condition affects only the LR part of the exchange correlation term, i.e., the LR parts get screened appropriately as $\beta$ changes from \( 1 - \alpha \) to  \( \frac{1}{\varepsilon} - \alpha \). With the constraints introduced so far, there is still no unique choice of $\alpha$. Different methods based on first principles considerations have been suggested to overcome this deficiency. One of the methods is based on a further property of the exact KS potential, namely the piecewise linearity of the total energy with respect to the (fractional) particle number. For example in Refs. \onlinecite{Srebro2012} and \onlinecite{Refaely-Abramson2012}, $\alpha$ was chosen such that the curvature, and therefore the deviation from linearity, of the total energy versus particle number curve is minimized. According to Stein et al., \cite{Stein2012} it may be possible to obtain the optimal $\alpha$ directly from minimization of the target function \( J^2(\gamma;\alpha) \), because deviations from piecewise linearity and from the IP theorem are two sides of the same coin. For cases where this fails to indicate a unique optimal value of $\alpha$, a more pragmatic approach which uses the good agreement of \emph{shifted} PBE0 results with experiment has been suggested.\cite{Egger2014} In that approach, $\alpha$ is determined so that the energy difference between the highest occupied delocalized state and the highest occupied localized state best correspond to a reference PBE0 calculation. 

\subsection{Computational details}

All geometry optimizations have been performed using the PBE functional. In order to circumvent issues concerning van-der-Waals interactions, which are poorly described in standard GGA and hybrid functionals,\cite{Sony2007,Romaner2009,Klimes2010,Kronik2014a} we employed the empirical correction scheme of Grimme \cite{Grimme2006} during the geometry optimization of the bulk structure. 
Note that we have taken lattice parameters from experiment\cite{Paulus2007,Panina2007} and only optimized the internal degrees of freedom.
The electronic structure of the isolated quinacridone molecule were obtained using QCHEM version 4.0 \cite{qchem4} with the cc-PVTZ basis set.\cite{Dunning1989}  DFT solid-state calculations of the crystal $\beta$-phase were performed using the PARATEC planewave package,\cite{Ihm1979} modified to include the new SRSH functional.\cite{Refaely-Abramson2013} Within PARATEC, GGA-based Troullier-Martins norm-conserving pseudopotentials\cite{Troullier1991} were employed \footnote{FHI-type pseudopotentials were adapted from the ABINIT website, \url{http://www.abinit.org/downloads/psp-links/psp-links/gga_fhi}, with core radii (in a.u.) of 1.276 for H, 1.498 for C, 1.399 for O, and 1.416 for N} to represent the core electrons and nuclei. We used a Monkhorst-Pack grid of $3 \times 3 \times 2$    $k$-points \cite{Monkhorst1976}.  

Perturbative G$_{0}$W$_{0}$ results, including the DFT results that serve as their starting point, were obtained using the VASP package,\cite{Kresse1993,Kresse1999,Shishkin2006} with both the PBE \cite{Perdew1996} and the HSE \cite{Heyd2006} functionals. The projector augmented wave (PAW) \cite{Bloechl1994} approach was employed for treatment of the core electrons, allowing for a relatively low kinetic energy cut-off of about 400 eV. The same $3 \times 3 \times 2$ $k$ Monkhorst-Pack grid employed above was used. 

G$_{0}$W$_{0}$ calculations were performed with full-frequency integration, using 48 frequency grid points and approximately 4000 unoccupied states to obtain the dielectric function and the self-energy. The scalar dielectric constant was determined as $\frac{1}{3}$ of the trace of the dielectric tensor obtained within the random phase approximation, including local field effects.\cite{Gajdos2006}

\section{Results and Discussion}

\subsection{Gas phase of quinacridone}

\begin{figure*}[!htb]
	\includegraphics[width=18cm]{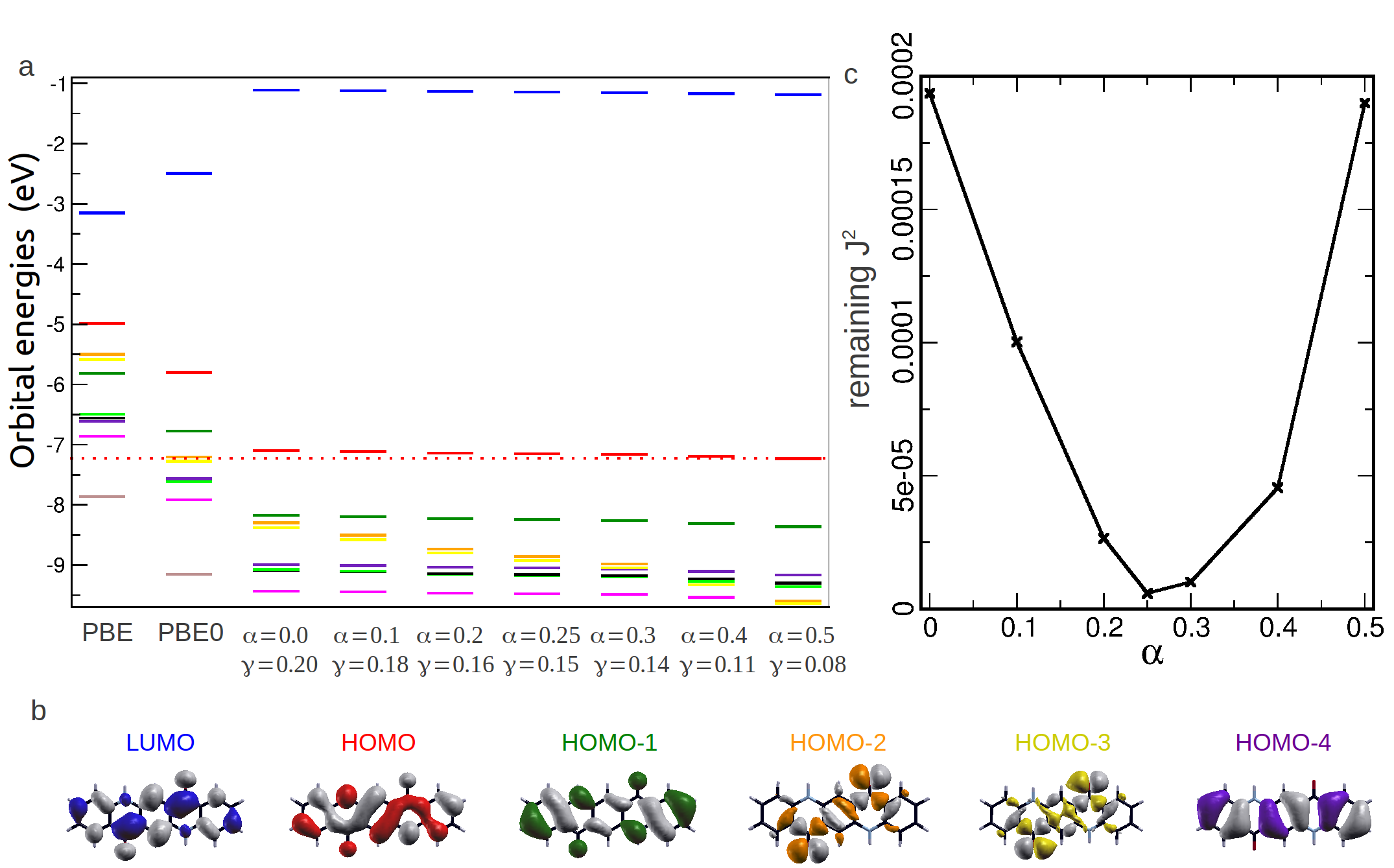}
	\caption{\label{Fig2} (a) DFT eigenvalue spectra of an isolated 5Q molecule, calculated with PBE, PBE0, and OT-RSH for for various values of the SR Fock-exchange parameter $\alpha$. For each choice of $\alpha$, the optimal range-separation parameter $\gamma$ is also denoted. The same color represents the same orbital in the different calculations. The red-dotted line denotes the experimental gas-phase UPS result for the IP.\cite{Slattery2001} (b) Images of the orbitals in the same color code, with the ordering resulting from the OT-RSH calculations with $\alpha=0.25$. (c)  Deviation from optimal tuning, \( J^2(\gamma;\alpha) \) as a function of $\alpha$, using the optimal value for $\gamma$ in each case.}
\end{figure*}

In this section, we present theoretical results for an isolated 5Q molecule, in order to characterize its electronic structure and frontier orbitals. Importantly, these calculations are crucial for obtaining the optimal $\alpha$ and $\gamma$ parameters, used in subsequent bulk calculations.
The optimal-tuning process, following Eq.~(\ref{eq:2}), has been determined using both \( i = 0 , 1 \) and \( i =  -1 , 0 , 1 \). Both calculations yielded similar results for the optimized values of $\gamma$, which deviated from each other by no more than 0.004 Bohr$^{-1}$. All results given here are the results of the \( i = 0 , 1 \) tuning.

Fig.~2a shows the eigenvalue spectra of an isolated 5Q molecule calculated with various xc functionals: PBE, PBE0, and OT-RSH results with different values for the parameter $\alpha$. For each value of $\alpha$ and the corresponding $\beta$=1-$\alpha$, the range-separation parameter $\gamma$ has been optimized separately. The PBE calculation results in a HOMO (red) value of 5.0 eV and a HOMO-LUMO (blue) gap of only 1.8 eV. The former is too low, by more than 2 eV, compared to the gas phase UPS IP value of 7.23 eV (shown as a red dotted line in Fig.~2).\cite{Slattery2001} In the PBE0 calculation, in which a 25\% of non-local Fock exchange is included, the HOMO level is 5.8 eV, still showing a large underestimation of the experimental reference, and the band gap increases to 3.3 eV. 

Next we consider the 5Q HOMO and LUMO values obtained with the OT-RSH approach. We have varied $\alpha$ from 0.0 to 0.5 in steps of 0.1 and optimized $\gamma$ for each $\alpha$ value. We observe that the HOMO value is improved to $\sim$ 7.15 eV, which slighlty increases for increasing $\alpha$ (within $\pm$ 0.05 eV), in excellent agreement with the experimental value of 7.23.\cite{Slattery2001} The HOMO-LUMO gap is increased to approximately 6.1 eV, and is only slightly ($\pm$ 0.05 eV) affected by the choice of the parameter $\alpha$. 

Turning to the outer-valence spectra, PBE yields an orbital ordering of the lower-lying occupied states that is different from that of all other calculations. In particular, the states shown as orange and yellow lines in Fig.\ref{Fig2}, i.e., the HOMO-1 and HOMO-2 of the PBE calculation, are located at higher energies relative to all other orbitals. The PBE0 results show a change in the eigenvalue spectra, along with energy-level "stretching".  In our OT-RSH results, the orbital ordering is parameter-dependent: all orbitals with $\pi$-symmetry (which clearly exhibit a similar degree of delocalization) show only little sensitivity to $\alpha$. In contrast, the $\sigma$-orbitals (HOMO-1 and HOMO-2 of the PBE calculation) are strongly affected and are shifted downwards by more than 1 eV when $\alpha$ changes from 0.0 to 0.5. This is not surprising: similar observations were made by Refaely-Abramson et al.\cite{Refaely-Abramson2012}  and confirmed in additional studies. \cite{Korzdorfer2012a,Egger2014,Koppen2014}

In order to provide an explanation for the origin of this behavior, orbital plots of the five highest-occupied orbitals, as well as the LUMO, are shown in Fig.~2b. Note that the color code for the isosurfaces is the same one used in the level diagrams of Fig.~2a. Comparing the shape of all orbitals in the probed energy range, one recognizes the $\sigma$ symmetry and the higher degree of localization of the two orbitals mentioned above. Building on experience with other organic molecules,\cite{Dori2006,Marom2008,Korzdorfer2009,Marom2009,Korzdorfer2010,Korzdorfer2010a,Ren2011a,Bisti2011,Refaely-Abramson2012} the reason for the different orbital energies and ordering between PBE and hybrid calculations is assigned to their different self interaction error (SIE). It should also be noted that these two $\sigma$-orbitals are the main difference between 5Q and pentacene, which has no $\sigma$-orbitals in the energy range of 5 eV below the HOMO.\cite{Berkebile2008,Korzdorfer2009,Korzdorfer2010} It has been previously shown\cite{Korzdorfer2010a} that all outer-valence frontier orbitals of pentacene exhibit similar SIE, and therefore orbital ordering in pentacene is less sensitive to the choice of the xc functional.

 Finally, we also observe that the optimized $\gamma$ parameter decreases with increasing amount of short-range Fock exchange. This can be rationalized by the range $\frac{1}{\gamma}$ at which full Fock exchange sets in, which can be extended to larger distances if the amount of Fock exchange at SR, governed by $\alpha$, is increased.\cite{Refaely-Abramson2012,Korzdorfer2012a,Egger2014} Fig.~2c shows the minimal $J^2$, obtained for the optimized $\gamma$ value for each $\alpha$, as a function of $\alpha$. The curve shows a distinct minimum of \( J^2(\gamma;\alpha) \) for $\alpha$ values between 0.2 and 0.3 (note the scale bar). It was shown\cite{Stein2012,Egger2014} that there is a rigorous quantitative equality between deviations from piecewise linearity and deviations from the IP theorem, represented by $J^2$.  We therefore chose the $\alpha$ value of 0.25, that minimizes $J^2$, to study the electronic structure of the bulk. This optimal value remains unchanged when including \( i =  -1 , 0 , 1 \) in the $\gamma$-tuning, or when comparing the energy difference between localized and delocalized states with PBE0,\cite{Egger2014} as discussed above.

\subsection{Solid $\beta$-phase of quinacridone}

\begin{figure}[!htb]
	\includegraphics[width=\columnwidth]{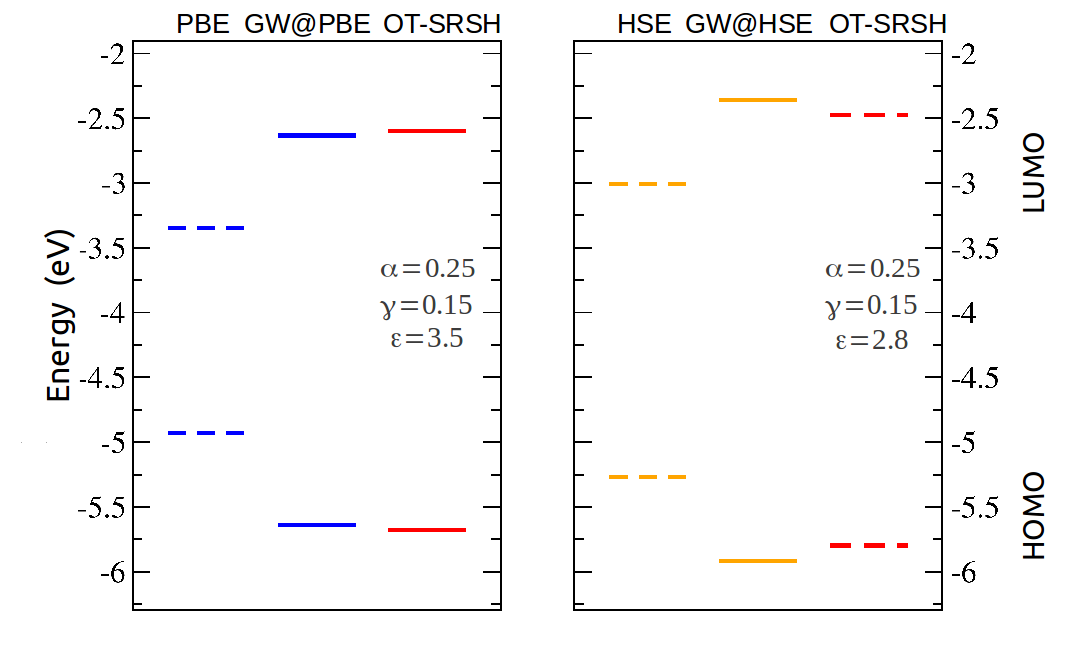}
	\caption{\label{Fig3} Band gaps of the 5Q $\beta$-crystalline structure obtained from different calculations. The peak maxima of the HOMO and LUMO derived bands in the corresponding density of states curves are shown as horizontal lines. In each case, the middle of the band gap is aligned with that of the computed gas-phase HOMO-LUMO gap.}
\end{figure}

We now turn our attention to 5Q in the solid $\beta$-phase. According to Eq.~(\ref{eq:4}), the scalar dielectric constant $\varepsilon$ governs the asymptotic behavior of the xc potential.  For organic molecular crystals, the short-range interactions are mainly governed by the molecule properties. We therefore take $\alpha=0.25$ and the optimized $\gamma=0.15$, as obtained from the above-discussed OT-RSH calculation for the isolated molecule. We take $\varepsilon$ to be 3.5, a value obtained from RPA calculations based on PBE eigenvalues, because it is already available as a by-product of our G$_0$W$_0$ results. Note, however, that it could easily be taken from computationally inexpensive approaches, \cite{Ruini2002,Romaner2008a,Natan2010,Schatschneider2013,Heitzer2013} and that this, in fact, is the recommended procedure if a comparison with GW is not performed. 

The results of the various DFT and GW calculations for the band-gap are summarized in Fig.~3. Note that due to the crystal structure exhibiting two molecules per unit cell, each molecular state splits into two bands in the crystal. In Fig.~3 we have defined the band gap as the peak-to-peak energy difference derived from the computed density of states. Also note that in the bulk calculation the absolute energy position of the highest-occupied and lowest-unoccupied levels are ill-defined, due to the lack of a reference vacuum energy. Therefore, we have aligned the center of the band gaps with those of the corresponding gas phase calculations. It must also be noted that, strictly speaking, a comparison of the computed "bulk" IPs and EAs with experimental values determined from thin films is problematic due to surface effects not being accounted for in the calculation.\cite{Duhm2008a,Sharifzadeh2012} 

When comparing the values for the fundamental gaps we find that, as expected, the PBE gap of about 1.6 eV is much smaller than those obtained with all other approaches. More importantly, it is also approximately unchanged compared to the isolated molecule, because no effects arising from the polarization of the environment are accounted for with the PBE xc functional.\cite{Neaton2006,Refaely-Abramson2013}
When computing G$_{0}$W$_{0}$ corrections to the PBE eigenvalues (GW@PBE), the gap increases to 3 eV. Note that polarization effects are inherently taken into account in these results as they are contained in the self-energy expression, leading to a considerably smaller gap than for the isolated molecule.\cite{Neaton2006} 
With our OT-SRSH calculation, we obtained a band gap of 3.1 eV, which is essentially the same as for GW@PBE. Thus the bulk band gap is roughly halved, compared to the corresponding calculation of the HOMO-LUMO gap in the isolated molecule.  

While for the isolated molecule the choice of $\alpha$ barely influences the band gap, for the bulk we observe a slightly different behavior. When reducing $\alpha$ from 0.25 to 0.0, but keeping  $\varepsilon=3.5$ unchanged, we observe a 0.3 eV reduction of the band gap. 
This finding can be explained by considering Eq.~(\ref{eq:4}). When reducing $\alpha$, the optimized $\gamma$ increases. In other words, when decreasing the amount of SR Fock exchange, the amount of LR Fock exchange increases to maintain the amount of overall non-local exchange. As a consequence, the spatial LR region in which the effective dielectric screening acts extends, thereby enhancing polarization effects and leading to a smaller band gap.

Now we compare the PBE results with those based on a HSE calculation. HSE yields a band gap of 2.3 eV, which lies between the pure PBE and the corresponding OT-SRSH result. In order to provide an explanation, recall the properties of the HSE functional. As already mentioned, it is a SR hybrid functional using non-local exchange only in the SR and pure semi-local exchange in the LR. The amount of Fock exchange is given by $\alpha = 0.25$ and a universal $\gamma$ value of 0.11 is used. Thus, the result is improved over PBE as some non-local exchange is introduced. However, because the xc potential decays exponentially, i.e., $\varepsilon = \infty$ in Eq.~(\ref{eq:4}), the asymptotic behavior is incorrect and a smaller band gap than in the OT-SRSH calculation is obtained.\cite{Paier2006,Refaely-Abramson2013} Compared to PBE, the increased HSE band gap results in the fact that a subsequent RPA calculation yields a decreased scalar dielectric constant of $\varepsilon = 2.8$. When using this value in the OT-SRSH calculation, we obtain a band gap of 3.3 eV, which is larger compared to the OT-SRSH band gap achieved using the PBE-based RPA $\varepsilon$ due to the reduced screening with the smaller HSE-based $\varepsilon$. Again the band gap compares well with a G$_{0}$W$_{0}$ computation with an HSE starting point (GW@HSE), which yields a band gap of 3.5 eV. The SRSH gaps are then indeed consistent with the GW calculation, given a similar scalar dielectric constant.

\begin{figure}[!htb]
	\includegraphics[width=\columnwidth]{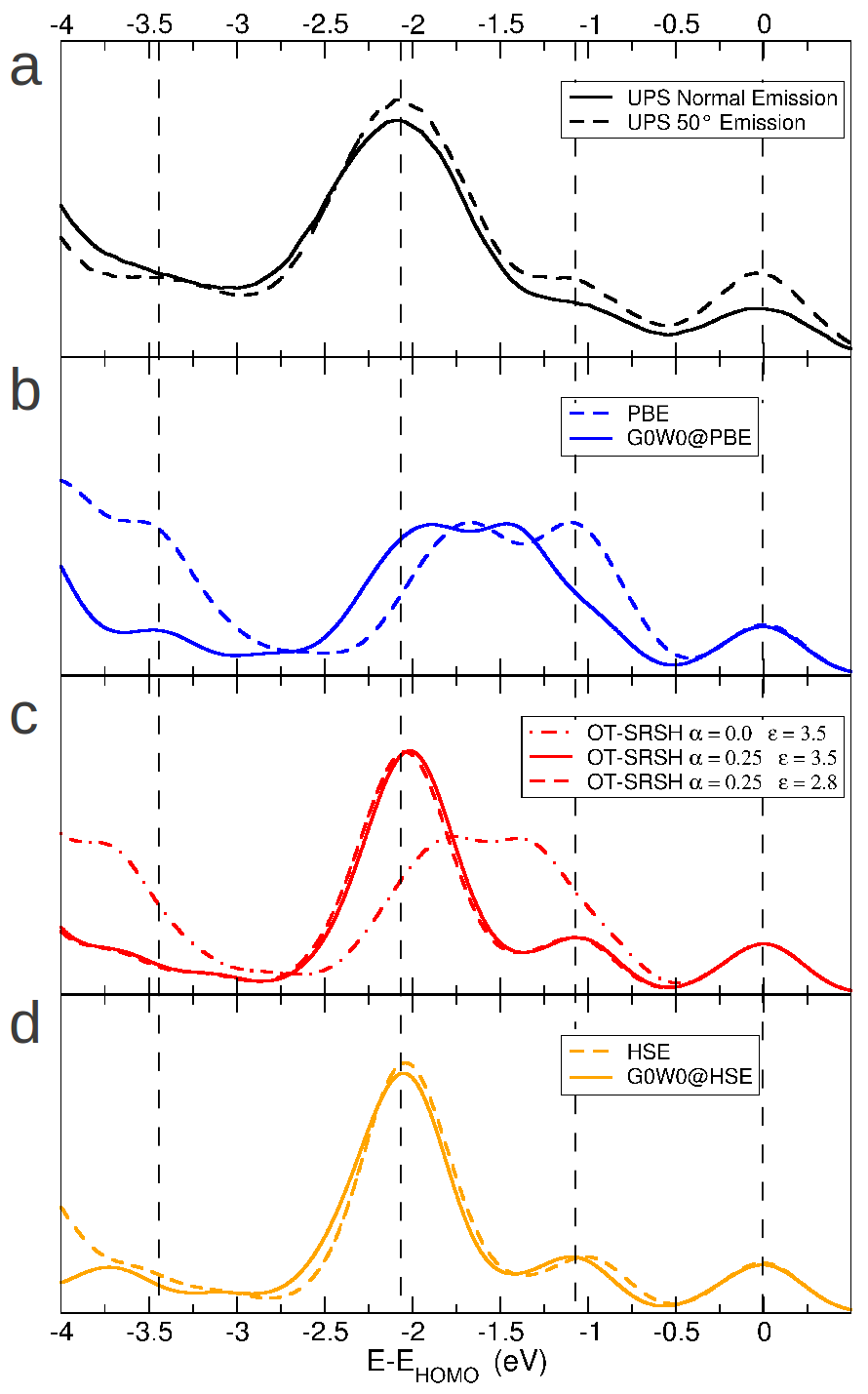}
	\caption{\label{Fig4} Comparison of experimental and theoretical photoemission spectra for the $\beta$-phase crystalline structure of 5Q. (a) Experimental UPS data in normal- and 50$^\circ$ emission (black solid and dashed lines, respectively). (b) Computed results of PBE  and GW@PBE  (blue dashed and solid lines,respectively). (c) Computed results of OT-SRSH approach with $\alpha = 0.0$ and $\alpha = 0.2$ (red dashed and solid lines, respectively). (d) Computed results of HSE and GW@HSE (orange dashed and solid line, respectively). For all spectra shown the energy axis has been aligned with the maximum of the highest occupied peak, which has been set to zero. A Gaussian broadening of 0.2 eV was used in all computed spectra.}
\end{figure}

Next, we investigate the valence band electronic density of states of the crystal, calculated at various levels of DFT and G$_{0}$W$_{0}$, in more detail. These are compared with experimental ARUPS results. The results are summarized in Fig.~4. Note that both experimental and calculated spectra have the energy axis shifted such that the highest occupied peak maximum has been aligned to zero. A Gaussian broadening of 0.2 eV was used in all computed spectra.

In the experimental spectra, there are four peaks in the shown energy range, which are highlighted by vertical dashed black lines. The first peak coincides with 0 eV, by virtue of the alignment procedure. The other peaks are located at $-1.1$, $-2.0$ and $-3.4$ eV.
For further interpretation of the experimental findings and for gaining more insight into the origin of individual peaks, we compare the peak positions and spectral shape of the experimental data to the theoretical results. In Fig.~\ref{Fig4}b, we display the PBE result (blue, dashed line) and the corresponding GW@PBE calculation (blue, solid line). 
At first sight, agreement between the PBE result and experiment appears reasonable because the peak positions seemingly agree quite well. However, when taking into account the peak shape as well, this agreement turns out to be coincidental. As in the case of the isolated molecule, PBE provides the wrong orbital energies and ordering, particularly for the localized states of $\sigma$ symmetry. 
The GW@PBE results certainly improve the band gap, as shown in Fig.~\ref{Fig3}, and also somewhat stretch the valence band spectrum. However, judging by the comparison with the experimental data, the GW@PBE result does not seem to accurately describe the investigated system. The main deviation of the GW@PBE curve is its peak at $-1.4$ eV, which does not show up in the experimental data at all. 

\begin{figure}[!htb]
	\includegraphics[width=\columnwidth]{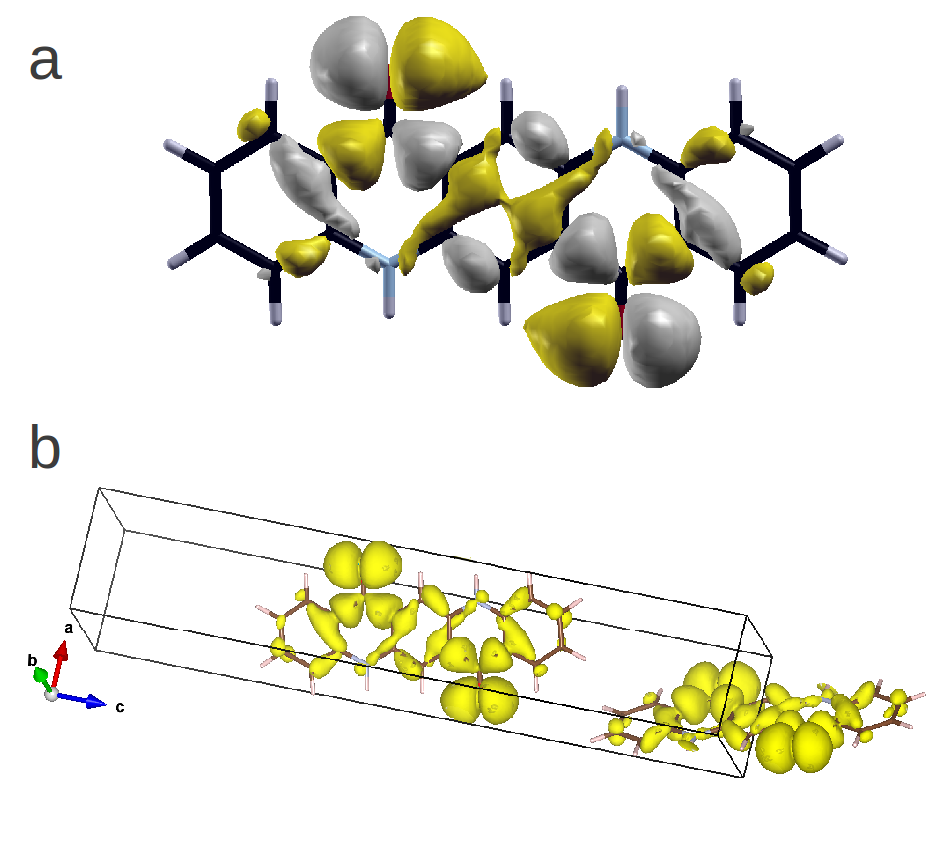}
	\caption{\label{Fig5} (a) Density plot of the HOMO-3 orbital obtained from the OT-RSH calculation of the isolated molecule. (b) Partial charge density decomposed on the band derived from the orbital shown in (a).}
\end{figure}

In order to clarify this deficiency of the GW@PBE result, we compute the valence band DOS resulting from the OT-SRSH approach using $\alpha = 0.25$, shown as a red solid line in Fig.~4c. From the previous findings for the isolated 5Q molecule, as well as similar molecules,\cite{Refaely-Abramson2012} we know that the amount of SR Fock exchange mainly affects states with a distinct degree of localization. Such behavior is also expected for the bulk. Therefore, Fig.~\ref{Fig4}c shows in addition the DOS obtained from an OT-SRSH calculation with $\alpha = 0.0$ as a red dash-dotted line. Indeed the two mentioned spectra in Fig.~\ref{Fig4}c are dramatically different, although only little influence of $\alpha$ on the size of the band gap was observed for the molecule (Fig.~\ref{Fig2}) and the molecular crystal, as discussed above. Interestingly, the $\alpha=0$ curve resembles the GW@PBE result, including a peak at about $-1.4$ eV, while the $\alpha=0.25$ spectrum has no peak at that energy, and the corresponding states are shifted to lower energies. This enhances the peak at $-2$ eV and leads to a rather impressive agreement with the experimental data. 
Thus, the origin of the incorrect peak at $-1.4$ eV in the GW@PBE and the OT-SRSH with $\alpha=0$ is related to a remaining SIE of the strongly localized $\sigma$ states (depicted in yellow and orange in Fig.~\ref{Fig2}), resulting with calculated energy levels that are too high for these states and changing the overall spectral shape. This is confirmed by plots of the orbital density associated with these states. In Fig.~\ref{Fig5}, a density plot of the HOMO-3 orbital of the isolated molecule obtained from OT-RSH with $\alpha = 0.25$, as well as the partial charge-density of the corresponding orbital in the solid-state, is shown.\footnote{Note that in the solid state the orbital ordering of states close in energy is dependent on a particular k-point, due to the band dispersion.} The extended bulk-state can be clearly associated with the respective orbital of the isolated molecule and the same is true for all other outer valence states (not shown for brevity).

Finally, we compare these results with those based on a HSE calculation. The spectra of the HSE and the corresponding GW@HSE calculation are shown in Fig.~\ref{Fig4}d as orange dashed and solid lines, respectively. Moreover, the OT-SRSH spectra with $\alpha=0.25$ and using $\varepsilon = 2.8$ from the HSE-based RPA calculation is displayed in Fig.~4c as a red dashed line. Clearly the two SRSH calculations with $\alpha=0.25$ coincide, up to a very small deviation. This observation shows that the screening introduced in Eq.~(\ref{eq:4}) affects all occupied states similarly and shifts the whole spectrum rigidly, thereby changing the computed band gap appropriately. Namely, while the $\varepsilon$ value greatly affects the gap renormalization (as it is the measure of electrostatic polarization), the shifted occupied spectra is almost entirely dependent on the values of  $\alpha$ and $\gamma$, and is practically the same for the two examined values of $\varepsilon$. The line shape of the outer-valence band spectrum of the HSE calculation is in good agreement with experiment and OT-SRSH calculations ($\alpha = 0.25$). The G$_{0}$W$_{0}$ calculation with the HSE starting point shift the HSE spectrum rigidly and we find an almost perfect agreement with the experiment. Furthermore, it becomes obvious that the full-frequency GW@PBE calculation suffers from the inappropriate starting point, as already reported in a number of studies of the organic and metal-organic molecule. \cite{Marom2011,Korzdorfer2012,Salomon2013}

\begin{figure}[!htb]
	\includegraphics[width=\columnwidth]{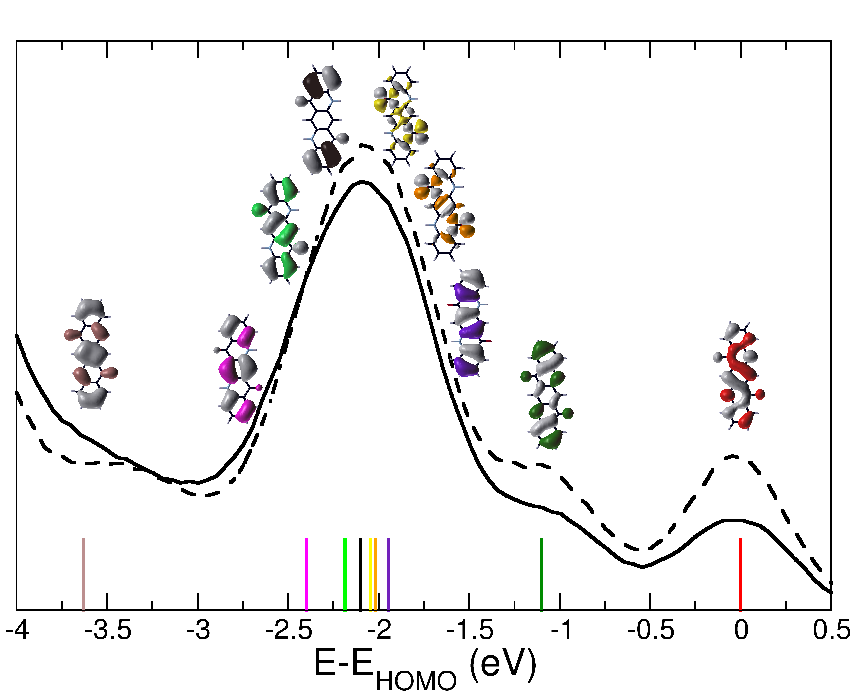}
	\caption{\label{Fig6} Measured (normal and 50$^\circ$) photoemission spectra for the $\beta$-crystalline structure of 5Q, including density plots of the contributing molecular states, as well as the energy ordering shown as vertival lines, obtained using the OT-SRSH functional with $\alpha = 0.25$, $\gamma=0.15$ and $\varepsilon = 2.8$.}
\end{figure}

Having found the theoretical methods which yield an accurate description of the electronic structure of the organic molecular crystal, we are able to assign specific molecular states to the experimentally observed peaks (a band-decomposed charge-density plot of one of these states was given in Fig.~\ref{Fig5}b). In Fig.~\ref{Fig6} we follow such plots to assign molecular orbitals to the appropriate peak in the OT-SRSH functional calculation with $\alpha = 0.25$. This optimal fraction of SR Fock exchange allows a simultaneous prediction of both $\sigma$-type localized orbitals (yellow and orange in Fig.~\ref{Fig6}) and $\pi$-type delocalized orbitals (other orbitals in Fig.~\ref{Fig6}), as discussed above. By that, it allows the assignment of theoretical orbitals to peaks of the experimental UPS data, as shown in Fig.~\ref{Fig6}.  

%\begin{figure}[!htb]
%	\includegraphics[width=\columnwidth]{Fig6.png}
%	\caption{\label{Fig6} Calculated density of states using a OT-SRSH functional with $\alpha = 0.25$ and $\varepsilon = 2.8$ including density plots of the contributing molecular states as well as the energy position of the peak maximum of corresponding bands.}
%\end{figure}  

\section{Conclusion}
 
In conclusion, we have studied the electronic structure of the organic molecule quinacridone in the gas phase and in the crystalline $\beta$-phase. 
For the gas phase, we find that the semilocal PBE and the hybrid PBE0 functional both underestimate the ionization potential and the fundamental band gap. When employing an optimally-tuned range separated hybrid functional, we obtain an excellent agreement with the experimental UPS value for the ionization potential and a larger fundamental gap, which is insensitive to the fraction $\alpha$ of short-range Fock exchange. The latter parameter, on the other hand, is shown to be crucial for attaining the correct relative orbital ordering of delocalized versus localized orbitals.

For the crystalline bulk $\beta$-phase of quinacridone, we have conducted angle-resolved photoemission experiments, which were used to benchmark our calculations. We demonstrate that by using the optimally-tuned value for the fraction of short-range exchange ($\alpha=0.25$), the correct orbital ordering was obtained within the occupied manifold of the states. We further showed that in order to take into account the level renormalization due to electronic polarization in the crystalline phase, the appropriate asymptotic behavior of the exchange-correlation functional is essential. This has been achieved by employing the optimally-tuned \emph{screened} range-separated hybrid (OT-SRSH) approach, in which the screening is accounted for by an effective dielectric constant, $\varepsilon$, which we computed from the trace of the macroscopic dielectric tensor obtained within the random phase approximation. Thus, the description of the unoccupied manifold of the states, in particular the fundamental gap, is also greatly improved, showing a band gap renormalization from the gas phase to the bulk based on physical grounds.  For comparison, we computed the $G_0W_0$ corrected electronic structure of the bulk using both PBE-GGA and the short-range hybrid functional HSE. These results emphasize the importance of the starting point in this perturbative approach, where the $G_0W_0$@HSE essentially agree with the OT-SRSH approach.

In summary, our work shows that the reliability of the optimal-tuning approach for molecular systems can be extended to the valence spectrum of molecular solid systems and that results at a level of accuracy comparable with GW calculations can be achieved.  We emphasize that, based on physically motivated choices for the parameters, the OT-SRSH approach allows for an accurate description of the band gap and at the same time of the relative orbital energies of the outer valence spectrum without any empiricism. Therefore, it may serve as a computationally inexpensive and reliable tool.

\section*{Acknowledgements}

D.L. and P.P. acknowledge support from the Austrian Science Fund (FWF) project P23190-N16. We thank Eric Glowacki from the Institute for Oragnic Solar Cells (LIOS), Johannes Kepler University Linz for pointing out the interesting device properties of 5Q and supplying the molecule. D.L. acknowledges the hospitality of the Weizmann Institute of Science.
S.R.A. is supported by an Adams fellowship of the Israel Academy of Sciences and
Humanities. L.K. and S.R.A. acknowledge supported by the European Research Council, the Israel Science Foundation, the Germany-Israel Foundation, the Wolfson Foundation, the Hemlsley Foundation, and the Lise Meitner Minerva Center for Computational Chemistry.

%\section*{Supporting information}

%\bibliography{amdm}


\end{document}